\documentclass{svjour3}                     
\smartqed  
\usepackage{graphicx}
\usepackage{mathptmx}      
\usepackage{amsmath}
\begin{document}

\title{Data Analysis and Phase Ambiguity Removal in the ACES
Microwave Link}


\author{Lo\"\i c Duchayne \and Peter Wolf \and Luigi Cacciapuoti \and
Marc-Peter Hess \and Marco Siccardi \and Wolfgang Schafer }


\institute{Lo\"\i c Duchayne \at
              LNE-SYRTE, Observatoire de Paris, LNE, CNRS, UPMC\\
               61 av. de l'observatoire, 75014 Paris, France \\
              \email{loic.duchayne@obspm.fr}             \\
           \and
                Peter Wolf \at
              LNE-SYRTE, Observatoire de Paris, LNE, CNRS, UPMC\\
               61 av. de l'observatoire, 75014 Paris, France \\
                               Luigi Cacciapuoti \at
              ESTEC, European Space Agency \\
 Noordwijk, The Netherlands \\
                                Marc-Peter Hess \at
              Astrium Space Transportation \\
Friedrichshafen, Germany \\
                                Marco Siccardi \at
TimeTech GmbH, Stuttgart, Germany\\
                                Wolfgang Schafer \at
TimeTech GmbH, Stuttgart, Germany\\}

\date{Received: date / Accepted: date}

\maketitle

\begin{abstract}
The ACES (Atomic Clock Ensemble in Space) mission is an ESA - CNES
project with the aim of setting up onboard the International Space
Station (ISS) several highly stable atomic clocks with a microwave
communication link (MWL). The specifications of the MWL are to
perform ground to space time and frequency comparisons with a
stability of 0.3 ps at one ISS pass and 7 ps at one day. The ACES
mission has applications in several domains such as fundamental
physics, metrology or geodesy.

The raw measurements of the ACES MWL need to be related to the
scientific products including all corrections (relativity,
atmosphere, internal delays or phase ambiguities) and considering
all terms greater than  0.1 ps when maximized. In fact, the
mission aims at extracting physical variables (scientific
products) such as clock desynchronisation, electron content in the
ionosphere (TEC), or range (instantaneous distance between the
stations) from the code and phase measurements on ground and in
space and auxiliary data (orbitography, internal delays,
atmospheric parameters, ...).

To this purpose we have developed the complete model of the time
transfer at the required  0.1 ps level. We have then developed in
parallel two softwares: \begin{enumerate}
    \item a program to simulate
the raw MWL measurements which allows the addition of different
types of noise, biases, dead times in the measurements, phase
ambiguities, etc ...
    \item an algorithm which implements the MWL model
and provides the "scientific products" from the raw measurements.
\end{enumerate}

During the mission only the algorithm (2.) will be used, but the
program (1.) is necessary for testing purposes. The two softwares
are kept as independent as possible (different programming
languages, different algorithms ...) to ensure a maximum
efficiency of such tests. We provide some details on the software
and the tests, considering different cases from the simplest to
the more complex and realistic situation using real ISS
orbitography data and MWL measurement noise from the MWL
engineering model.

The phase ambiguity removal of carrier phase measurements is
performed by the algorithm and its success strongly depends on the
noise of the observables. We have investigated the statistics of
cycle slips which appear during this operation using experimental
data obtained from the tests of the MWL engineering model. We
present two novel methods which allow the reduction of the cycle
slip probabilities by a factor greater than 5 compared to the
standard method.
\end{abstract}

\section{ Introduction }

Due to recent scientific breakthroughs such as laser cooling and
atom trapping methods, huge progress has been achieved on the
uncertainties of atomic clocks during the last twenty years
\cite{Bize,Heavner, Oskay, Rosenband}. Some of them  reach a
precision of a few parts over $10^{17}$ in relative frequency.

In a Space environment these atomic sensors would become an
exceptional tool for promising applications in fundamental
physics, geodesy, time and frequency metrology or in navigation.
Onboard terrestrial or solar system satellites, their exceptional
properties allow them to test the fundamental laws of nature, to
study the Earth's and solar system gravitational potential and its
evolution, or to explore the Universe \cite{Sagas}.

To this purpose, the ACES (Atomic Clock Ensemble in Space) mission
\cite{Salomon}, an ESA-CNES project will be installed onboard the
ISS (International Space Station) in 2013. It consists of two
atomic clocks and a two-way time transfer system (microwave link,
MWL) with an overall uncertainty goal of 1 part in $10^{16}$ after
ten day integration (see section \ref{ACES} for more details).
This mission will  carry out tests of fundamental physics such as
testing the Einstein's Equivalence Principle or perform time
transfer at the sub-picosecond level.

This next generation of space clocks at the envisaged uncertainty
level requires a fully relativistic analysis, not only of the
clocks (in space and on the ground) but also of the time/frequency
transfer method used to compare them
\cite{Allan,Klioner,Petit,Wolf1,Blanchet,Teyssandier}. Similarly
the modeling of the mission for the future data analysis requires
a relativistic framework and the investigation of the complete
model complying with the envisaged performances.

The preparation of the mission data processing needs the
development of two softwares. On one hand, because the mission is
still in preparation, we need to simulate the MWL raw measurements
in the most realistic way possible. On the other hand, the mission
requires an algorithm to extract the "scientific products" from
the raw measurements and the parameters of the mission.

Finally, similarly to Global Positioning Systems, the phase
ambiguity resolution is a key process which the mission success
depends on. In fact an error of one cycle on one of the three
frequencies will lead to an error larger than the mission
specifications. The occurrence probability of cycle slips strongly
depends on the raw measurement noise and observable combinations
must be found to reduce this probability.

In this paper we study in more detail the preparation of the data
analysis and provide some details on its testing which will allow
to evaluate the high performances of the MWL. In sections
\ref{ACES} and \ref{MWLmodel} we briefly describe the ACES
mission, and then the relativistic model used for the clocks and
the time transfer to understand the nature of the raw
measurements. We also relate their expression with all propagation
effects between the two clocks (internal delays, atmospheric
effects, relativistic contributions,...). This investigation is
required to model the time transfer at the expected level. In
section \ref{DataAnalysis} we give information about the
measurement simulation and the scientific products extracting
algorithm and show some tests of the latter to evaluate its
performances in terms of stability and accuracy. In the two last
sections (\ref{PhaseAmbiguityResolution} and \ref{TestsNoise})
before the conclusion, we discuss the phase ambiguity
determination whose success strongly depends on the noise of the
observables. We investigate the statistics of cycle slips which
appear during this operation. To this purpose we use experimental
data obtained from the tests of the MWL engineering model
\cite{Schaefer}, and an ISS ephemeris.

%
%

\section{The ACES mission}
\label{ACES}

 The ACES project led by the CNES and the ESA aims at
setting up on the ISS several highly stable clocks around 2013.
The ACES payload includes two clocks, a hydrogen maser (SHM
developed by TEMEX) and a cold atom clock PHARAO (developed by
CNES) respectively for short and long term performances, and a
microwave link for communication and time/frequency comparison.
The frequency stability of PHARAO onboard the ISS is expected to
be better than $10^{-13}$ for one second, $3 \cdot 10^{-16}$ over
one day and $1 \cdot 10^{-16}$ over ten days, with an accuracy
goal of $1 \cdot 10^{-16}$ in relative frequency.

The ACES mission has as objectives :
\begin{itemize}
    \item to operate a cold atom clock in microgravity with a 100 mHz
    linewidth,
    \item to compare the high performances of the two atomic clocks in space
    (PHARAO and SHM) and to obtain a stability of $3\cdot 10^{-16}$
    over one day,
    \item to perform time comparisons between the two space clocks and ground clocks,
    \item to carry out tests of fundamental physics such as a gravitational redshift
    measurement and to search for a potential speed of light
    anisotropy and a possible drift of the fine structure constant
    $\alpha$.
    \item to perform precise measurements of the Total Electron Content (TEC) in the ionosphere, the
    tropospheric delay and the Newtonian potential.
 \end{itemize}

The time transfer is performed using a micro-wave two-way system,
called Micro-Wave Link (MWL). An additional frequency is added in
order to measure and correct the ionospheric delay at the required
level. It uses carriers of frequency 13.5, 14.7 et 2.25 GHz,
modulated by pseudo random codes respectively at $10^{8}$
$s^{-1}$, $10^{8}$ $s^{-1}$ and $10^{6}$ $s^{-1}$ chip rates.
Moreover it has four channels that allow four ground stations to
be compared with the ISS clock at the same time.

According to the mission specifications, the microwave link has to
synchronize two atomic clocks with a time stability of
$\leq$~0.3~ps over 300~s, $\leq$~7~ps over one day, and
$\leq$~23~ps over 10 days. The performance of this link is a key
issue since it will perform high precision time comparisons
without damaging the high performances of the clocks.

For our purposes we express the above requirements for the MWL in
a simplified form by the temporal Allan deviation
$(\sigma_x(\tau))$ :

\begin{equation}
\sigma_x (\tau) = 5.2 \cdot 10^{-12} \cdot \tau^{-\frac{1}{2}}
\label{BruitIntrinseque}
\end{equation}
for a single satellite pass over a ground station (for integration
times $\tau$ lower than $300$ s) and by

\begin{equation}\label{AttentesAuxTempsLongs}
\sigma_x (\tau) = 2.4 \cdot 10^{-14} \cdot \tau^{\frac{1}{2}}
\end{equation}
for longer integration times (for integration times $\tau$ greater
than $300$ s).


We take (\ref{BruitIntrinseque}) and (\ref{AttentesAuxTempsLongs})
as our upper limits for all comparisons with the test results of
the data processing algorithm in the following sections.

\section{The time transfer model}
\label{MWLmodel}

 Due to the outstanding performances of the clocks,
some  effects from fundamental physics have an impact on the clock
behaviors and on the signal propagation. In this situation, a
relativistic point of view must be considered to model the clocks
and the time transfer. The aim of this section is to describe the
relativistic model used for the ACES mission and the Micro-Wave
Link. It is necessary to express the raw measurements as functions
of the clock desynchronisation and of all effects which affect the
signal propagation (relativity, atmosphere, internal delays, phase
ambiguities, etc...) at the expected level.

In a general relativistic framework each clock produces its own
local proper time, in our case $\tau^g$ and $\tau^s$ for the
ground and space clocks respectively. In order to model signal
propagation between the ground and the space stations, we use a
non-rotating geocentric space-time coordinate system. Thus
$t=x_0/c$ is the geocentric coordinate time, $\overrightarrow{x} =
(x_1, x_2, x_3)$ are the spatial coordinates, where c is the speed
of light in vacuum  (c = 299792458 $m.s^{-1}$). We denote
$U(t,\overrightarrow{x})$ as the total Newtonian potential at the
coordinate time $t$ and the position $\overrightarrow{x}$ with the
convention that $U \geq 0$ \cite{Soffel2}. In these coordinates,
the metric is given by an approximate solution of Einstein's
equations  valid for low velocity and potential
($\frac{U}{c^{2}}<<1$ and $\frac{v^{2}}{c^{2}}<<1$):

\begin{equation}
ds^{2} = -(1 - \frac{2 U(t, \overrightarrow{x})}{c^2})c^2 dt^2 +(1
+ \frac{2 U(t, \overrightarrow{x})}{c^2})d\overrightarrow{x}^2,
\label{Metric}
\end{equation}
where higher order terms can be neglected for our purposes
\cite{Wolf1}.

In this system, each emission or reception event (at the antenna
phase center) is identified by its coordinate time $t_i$ (figure
\ref{fig:ACESTrajectoires}) and a coordinate time interval is
defined by $T_{ij}=t_{j}-t_{i}$. We define $\overrightarrow{x}_g$,
$\overrightarrow{v}_g$ and $\overrightarrow{a}_g$ respectively as
the position, the velocity and the acceleration of the ground
station, and $\overrightarrow{x}_s$, $\overrightarrow{v}_s$ and
$\overrightarrow{a}_s$ respectively as the position, the velocity
and the acceleration of the space station.
\begin{figure}
        \includegraphics[height=5cm]{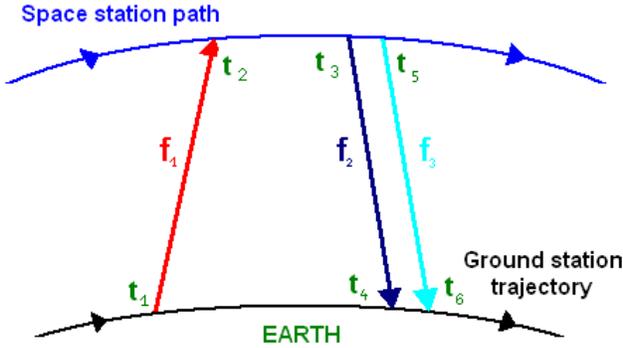}
    \caption{MWL principle}
    \label{fig:ACESTrajectoires}
\end{figure}


The $f_1$ frequency signal is emitted by the ground station at the
coordinate time $t_1$ and received by the space station at $t_2$.
The $f_2$ and $f_3$ frequency signals are emitted from the space
station at $t_3$ and $t_5$, and received at the ground station at
$t_4$ and $t_6$. The third frequency is added to measure the TEC
in the ionosphere which allows the correction of the ionospheric
delay.

According to equation (\ref{Metric}), we introduce the notation
$[.]^A$ to express an interval of the coordinate time $t$ in the
proper time scale $\tau^A$ produced by a clock at
$\overrightarrow{x}_A (t)$. The following equation relates a
coordinate time interval $T_{12}$ elapsed between the coordinate
time $t_1$ and $t_2$ to a proper time interval $[T_{12}]^A$
elapsed in the proper time scale $\tau^A$ :
\begin{equation}
[T_{12}]^A = \int_{t_1} ^{t_2}(1 - \frac{U (t,
\overrightarrow{x}_A)}{c^2} - \frac{v_A^{2}(t)}{2 c^2})dt,
\label{transforméentempspropredeA}
\end{equation}

where $\overrightarrow{x}_A$ and $v_A$ are respectively the
position and the norm of the velocity of a point A expressed in
the coordinate system described above, and the integral is
evaluated along the path of the clock $\overrightarrow{x}_A (t)$.

\ 

The MWL is characterized by its continuous way of emission. It
measures the time offsets between the locally generated signal and
the received one. It provides three measurements (or observables)
of the code (one on the space station, two on the ground) and
three measurements of the phase of the carrier frequency at a
sampling rate of one Hertz.

An observable is related to the phase comparison between a signal
derived from the local oscillator and the received signal,
corrected for the frequency difference mainly due to the first
order Doppler effect (see \cite{Bahder} for details of a similar
procedure used in GPS). If we consider a special bit of the signal
which is produced locally at $\tau_p$ and received at $\tau_a$, an
observable is a measurement of the local proper time interval
between these two events. The observable is labeled with the
arrival proper time $\tau_a$.

For instance, the upwards signal observable is
\begin{equation}\label{}
    \Delta \tau^{s} (\tau_a^{s}) = \tau_p^{s} -\tau_a^{s} = \tau_p^{g}
    -\tau_a^{s}.
\end{equation}

The last identity comes from the fact that two identical segments
of code are produced at the same local time :
$\tau_p^{g}=\tau_p^{s}$.

Thus we have
    \begin{equation}\label{}
    \Delta \tau^{s} \bigl(\tau^{s}(t_2)\bigl) = \tau^{g}(t_1)
    -\tau^{s}(t_2).
\end{equation}

We define the desynchronisation between the two clocks at the
coordinate time $t$ by the difference between the ground and the
space proper times at $t$, $\tau^{g}(t)
    -\tau^{s}(t)$. The raw
measurement can be expressed as a function of the
desynchronisation between the two clocks at the time of reception
$t_2$, the emission and reception internal delays, and the
propagation time between the two stations $T_{12}$. For the sake
of clarity and equation simplicity, we intentionally omit internal
delays in the following expressions. They can be associated with
the propagation time except the fact they are proper time
intervals in the local clock time scale, and their inclusion in
the model poses no particular difficulties.

For example, if we consider the observable from $f_1$ signal PRN
code, it is given by

    \begin{equation}\label{CodeC1}
    \Delta \tau^{s} \bigl(\tau^{s}(t_2)\bigl) = \tau^{g}(t_2)
    -\tau^{s}(t_2) - [T_{12}]^g.
\end{equation}
where $T_{12}$ is the propagation time for the $f_1$ signal. The
latter is a coordinate time interval transformed to a ground
proper time interval following  equation
(\ref{transforméentempspropredeA}).

As we want to evaluate the desynchronisation between the two
clocks, the time interval $T_{12}$ elapsed between emission from
the ground station and reception by the satellite of the $f_1$
frequency signal needs to be calculated. It can be written as

\begin{equation}
\begin{split}
T_{12} &= \frac{R_{12}}{c} + \frac{2 G
M_{E}}{c^{3}}\ln\left(\frac{x_{g}(t_1) + x_{s}(t_2) +
R_{12}}{x_{g}(t_1) +x_{s}(t_2) - R_{12}}\right)\\& +
\Delta_{12}^{tropo}+ \Delta_{12}^{iono}+O(\frac{1}{c^4}), \\
\label{ExpressionT12}
\end{split}
\end{equation}
where $R_{12} = ||\overrightarrow{R_{12}}|| =
||\overrightarrow{x_s}(t_2)-\overrightarrow{x_g}(t_1)||$, where
the logarithmic term represents the Shapiro time delay
\cite{Shapiro} (see e.g. \cite{Blanchet} for a detailed
derivation) and where $\Delta_{12}^{tropo}$ and
$\Delta_{12}^{iono}$ are respectively the tropospheric and
ionospheric delays on the signal path.

The phase observable for the $f_1$ signal can be expressed
similarly except for two important features : the effect of
ionosphere and the phase ambiguity. The ionosphere delay takes
opposite signs for code and phase \cite{Bassiri}. Moreover a
carrier phase measurement is less noisy than code measurement but
affected by a phase ambiguity. This ambiguity corresponds to
integer number $N_i$ of the signal period $1/f_i$. For instance
the phase measurement of the $f_1$ signal is given by

    \begin{equation}\label{PhasePhi1}
    \Delta \tau_\phi^{s} \bigl(\tau^{s}(t_2)\bigl) = \tau^{g}(t_2)
    -\tau^{s}(t_2)
    -[T_{12}^{\phi}]^g + N_1/f_1.
\end{equation}

where $T_{12}^{\phi}$ is the phase propagation time of the $f_1$
signal and given by (\ref{ExpressionT12}) with a change of sign of
the ionospheric term $\Delta_{12}^{iono}$.

An error in the determination of $N_i$ is referred to as a "cycle
slip". It will be studied in detail in sections
(\ref{PhaseAmbiguityResolution}) and (\ref{TestsNoise}).

The expressions of code and phase observables for the two
remaining frequencies can be derived similarly to (\ref{CodeC1})
and (\ref{PhasePhi1}). The combination of $f_1$ and $f_2$ signal
observables gives the expression of the desynchronisation between
ground and space clocks :

\begin{equation}
\begin{split}
\tau^g(t_a) - \tau^s(t_a)
 &=\frac{1}{2}\biggl(\Delta\tau^s\left(\tau^s(t_2)\right) - \Delta
\tau^g\left(\tau^g(t_4)\right) \\ &+ T_{12} - T_{34}  \\
 &- \int_{t_1}^{t_2}(\frac{U (t, \overrightarrow{x_g})}{c^2} +
\frac{v_g^{2}(t)}{2 c^2})dt \\
& + \int_{t_3} ^{t_4}(\frac{U (t,
\overrightarrow{x_s})}{c^2} + \frac{v_s^{2}(t)}{2 c^2})dt \biggr),\\
\label{ExpressionDesynMWLReceptionTime}
\end{split}
\end{equation}

where $t_a \equiv \frac{t_2 + t_4}{2}$, and where
$\Delta\tau^s\left(\tau^s(t_2)\right)$ and $\Delta
\tau^g\left(\tau^g(t_4)\right)$ are the observables respectively
from the ground and onboard the satellite, and where we have
neglected non-linearities of $\tau^g(t)$ and $\tau^s(t)$ over the
interval $t_4 - t_2$ (few milliseconds). The integral terms result
from proper time to coordinate time transformations. They are
small corrections of order $10^{-12}$ s to the desynchronisation
$\tau^g(t_a) - \tau^s(t_a)$.

In (\ref{ExpressionDesynMWLReceptionTime}), the difference
$T_{12}-T_{34}$ needs to be calculated from the knowledge of
satellite and ground positions and velocities obtained from orbit
restitution (see \cite{Duchayne}). It includes the determination
of ionospheric delays through the evaluation of the Total Electron
Content (TEC). This is done by combining the observables coming
from $f_2$ and $f_3$ signals because the signal paths are
approximatively the same and there is almost one order of
magnitude between the two frequencies which provides a good
estimation of the TEC.

The same equation exists for phase observables. Considering phase
ambiguities are correctly removed, it differs from the previous
equation (\ref{ExpressionDesynMWLReceptionTime}) by the term
$T_{12} - T_{34}$. This term is not the same for code and phase
desynchronisation equation due to the different effects from
ionosphere on signal propagation.

\section{ Data analysis and simulation}
\label{DataAnalysis}

The data analysis of ACES measurements has specific goals. It aims
at providing the searched scientific products from the MWL
measurements and some additional inputs, in particular ISS
orbitography, parameters for the troposphere model, etc... Its
goal is to evaluate physical parameters which will serve to study
the Earth's atmosphere or to probe the fundamental laws of
physics.

However, the data analysis has to deal with experimental issues
such as noise on measurements or loss of signal acquisition which
leads to dead times in the observables. Even in adverse
conditions, it has to provide results within the expected
performances of the mission. Thus the data analysis must be
prepared before the launch of the mission, and tested in all
pessimistic situations.

To this purpose, two software packages have been developed
considering the complete model of time transfer described in
\cite{Duchayne}. On one hand, a raw MWL measurement simulation
which produces the observables from given clock difference, orbit,
internal delays, ionosphere and troposphere parameters, and on the
other hand, a data analysis algorithm which extracts the searched
parameters from the raw measurements. The two softwares have been
implemented in the aim of keeping them as independent as possible.
That is why they use different languages of programmation and
different algorithms of calculation.

In the following paragraphs we will give some details on both
softwares.

\subsection{The raw measurement simulation}

The simulation produces the raw measurements of the MWL for code
and carrier phase on the three frequencies. To this purpose it
needs several parameters chosen by the user such as the clock
behaviors ($\tau^g (t)$ and $\tau^s (t)$) , the space and ground
station trajectories or the evolution of internal delays.

It is implemented in an object oriented language and simulates the
six observables using an iteration procedure.

As an example,  the simulation of the $f_1$ signal code
measurement $\Delta \tau^{s} \bigl(\tau^{s}(t_2)\bigl)$ measured
at the space proper time $\tau^s(t_2)$ starts with the calculation
of the propagation time $T$ between the ground and space stations
respectively located at $\overrightarrow{x_g}(t_2)$ and
$\overrightarrow{x_s}(t_2)$, which are obtained by interpolation
of the orbitography data. Due to the propagation delay, the ground
station was not at $\overrightarrow{x_g}(t_2)$, but at a previous
position along its trajectory when emitting the signal. New
position, velocity and acceleration are interpolated at the
coordinate time $t_2 - T$ which allows to calculate a new value
for the propagation time $T$. By convergence, this method provides
the true time of signal emission $t_1$ by alternatively
calculating the propagation time $T$ for a signal arriving at the
coordinate time $t_2$ and interpolating the ground station
position, velocity and acceleration. We then evaluate the proper
time $\tau^g(t_1)$ of signal emission from the ground station and
finally the code observable $\Delta \tau^{s}
\bigl(\tau^{s}(t_2)\bigl) = \tau^g(t_1) - \tau^s(t_2)$.

Once the "theoretical" measurement is calculated, perturbations
are added to it.  The simulation allows addition of measurement
noise, clock error, internal delays and errors on them, antenna
phase patterns, dead times in measurements, orbitography error,
etc ...

\subsection{The data processing algorithm }

The MWL data analysis implements the MWL model and provides the
scientific products from the raw MWL measurements. Its outputs are
the Total Electron Content (TEC) in the ionosphere along the line
of sight, the desynchronisation between the ground and the space
clocks, the tropospheric delay and the instantaneous distance
between the two stations. These products are obtained from code
only, and code plus phase measurements. Besides the raw
measurements, the calculation needs some other data such as the
orbitography of the ground and space stations, atmospheric
parameters (temperature, pressure, ...) or the internal delays.

Contrary to the raw measurement simulation, the data processing
algorithm is written in an assembly language. Before the
calculation of the scientific products, it performs a
preprocessing so as to detect dead times in code and phase
measurements, to combine correctly the measurements of the three
frequencies and to transform space and ground station trajectories
in the adapted coordinate system. Moreover it also determines
cycle ambiguities on the carrier phase measurements before using
the non-ambiguous phase observables to estimate the scientific
products.

The calculation is based on analytical expressions at a chosen
time. For instance, in expression
(\ref{ExpressionDesynMWLReceptionTime}),  the term $T_{12}-T_{34}$
is evaluated with a Taylor expansion. Its expression is a function
of position, velocity and acceleration of the two stations at one
chosen coordinate time. The same kind of expansions are used for
the calculation of TEC, range or tropospheric delay.

\subsection{Tests and results}

The testing of the algorithm is performed by taking into account
independently and then simultaneously the different noises and
perturbations which will appear during the mission (code and phase
measurements noise, clock noise, orbit restitution noise, dead
times, ...). It allows to verify its stability and accuracy
performances.

To this purpose we feed the output of the raw measurement
simulation into the data analysis algorithm and compare if its
output corresponds to the initial given functions used as input
for the simulation. The differences between the outputs of the
algorithm and the initial functions  are identified as final
errors on scientific products. In the actual mission, only the MWL
data analysis will be used, to obtain directly the required
scientific products.

\begin{figure}
        \includegraphics[height=5cm]{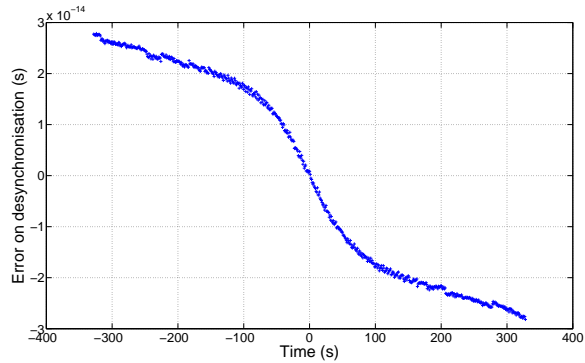}
    \caption{Error on desynchronisation when no noise is added}
    \label{fig:ATest1Desyn1Code}
\end{figure}

\begin{figure}
        \includegraphics[height=5cm]{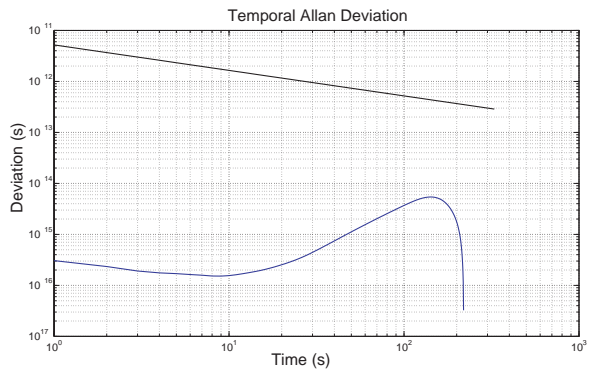}
    \caption{Temporal Allan deviation when no noise is added}
    \label{fig:TemporalAllanDeviationCas1}
\end{figure}

Many tests have been carried out, but for clarity, we only show
the results of three of them. First we add no noise which would
damage the performances of the link. The obtained error on
desynchronisation is drawn on figure \ref{fig:ATest1Desyn1Code}
and stays under $0.1$ picoseconds. The remaining term corresponds
to difference of tropospheric and ionospheric delays for Ku
signals ($f_1$ and $f_2$) which have not crossed the same
atmospheric layers. The resulting deviation of the error on the
desynchronisation is two orders of magnitude below the
specifications (cf. figure \ref{fig:TemporalAllanDeviationCas1}).
We conclude that no term giving a deviation over the
specifications has been neglected.

\begin{figure}
        \includegraphics[height=5cm]{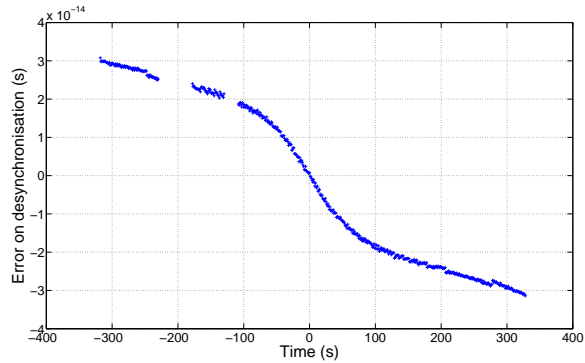}
    \caption{Error on desynchronisation with dead times in measurements}
    \label{fig:Test8Desyn1Code}
\end{figure}

Now the three observables have dead time intervals. They can be
caused for example by a loss of station visibility or to the
occultation of the reception antenna by a solar panel.
 Figure \ref{fig:Test8Desyn1Code} shows that the software
deals with dead times and provides the scientific products without
error. The corresponding deviation is also two orders of magnitude
under the specifications of the mission.

Finally the last presented test is performed adding white noise on
all simulated raw measurements with an amplitude corresponding to
the specifications (\ref{BruitIntrinseque}). The resulting
deviation stays under the specifications (cf. figure
\ref{fig:ACESTrajectoires}). It is calculated in accordance with
equation (\ref{ExpressionDesynMWLReceptionTime}) : the white
noises from $f_1$ and $f_2$ observables are not correlated which
explains the square root of two gain with respect to the mission
specifications. The addition of noise is handled by the data
analysis software and this is demonstrating the robustness of the
algorithm confronting this difficulty.

\begin{figure}
        \includegraphics[height=5cm]{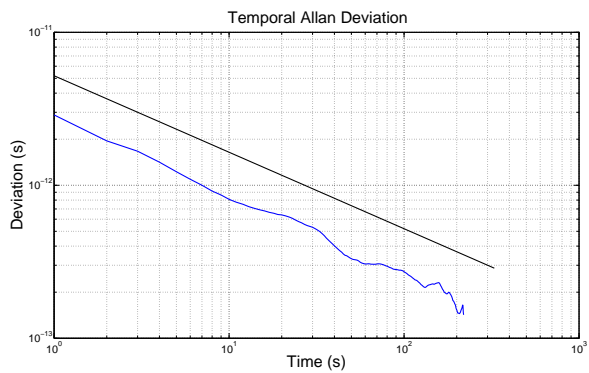}
    \caption{Temporal Allan deviation considering white measurement noise on observables}
    \label{fig:ACESTrajectoires}
\end{figure}

As written before, many other successful tests have been done
considering different perturbations affecting the accuracy and the
stability of the ACES time transfer. They involve realistic
orbitography error, random walk noise on clocks or offset and
noise on internal delays. The algorithm survived all these tests
which complies with its initial objectives. In addition, the
combination of the two softwares allows one to easily evaluate the
effects of each parameter on the time and frequency transfer or to
find out what the final results would be if there was an Einstein
Equivalence Principe violation.

\section{ Phase ambiguity resolution}
\label{PhaseAmbiguityResolution}

For all time/frequency transfer techniques the carrier phase
observables have the advantage of significantly  better precision
than  the code measurements, and therefore their use has been
implemented in most techniques. Using the phase observables
requires resolution of the phase ambiguities, which has been
investigated in detail in many techniques, in particular the
Global Positioning System \cite{Euler,Jonge}.

However there are several differences between GPS and the
MicroWave Link. On one hand, the MWL is a two way time transfer
with three frequencies involved. The combination of measurements
coming from an upwards and a downwards signal allows to eliminate
restraining factors without canceling the clock behaviors. On the
other hand, the frequencies used in the MWL are higher than GPS
frequencies except for the third MWL frequency which is added to
evaluate the TEC in the ionosphere. There is almost one order of
magnitude between the lowest and the highest frequencies. The
phase ambiguity identification must therefore be investigated for
the particular case of the ACES mission.

Similarly to equation (\ref{ExpressionDesynMWLReceptionTime}), the
 desynchronsation between the two clocks can be expressed as a function
of phase observables and cycle ambiguities. It allows to study the
consequences of the phase ambiguity detection on the final
calculation.

\begin{equation}
\begin{split}
\tau^g(t_a) - \tau^s(t_a) & =
 \frac{1}{2}\Biggl(\Delta\tau_\phi^s\left(\tau^s(t_2)\right) - \Delta
\tau_\phi^g\left(\tau^g(t_4)\right)
-\frac{N_1}{f_1}\\&+\frac{N_2}{f_2} + \Delta   - (\frac{1}{f_1^2}
- \frac{1}{f_2^2})\frac{f_2^2 f_3^2}{f_2^2-f_3^2} \times\\&
\biggl(\Delta\tau_\phi^s\left(\tau^s(t_4)\right)  - \Delta
\tau_\phi^g\left(\tau^g(t_6)\right) -\frac{N_2}{f_2}+\frac{N_3}{f_3}\biggr)   \Biggr),\\
\label{ExpressionDesynMWLReceptionTimePhase}
\end{split}
\end{equation}

where the last term is the correction for the leading order
ionospheric effect and $\Delta$ includes terms due to path
asymmetries, tropospheric delays, relativistic corrections, higher
order ionospheric terms, etc... (see \cite{Duchayne} for details).
Although these terms need to be taken into account, they are not
relevant for the discussion concerning phase ambiguity resolution
and will be neglected in the following.

According to equation
(\ref{ExpressionDesynMWLReceptionTimePhase}), the cycle slip
detection is necessary for the three frequencies. In fact an error
of one cycle on $f_1$, $f_2$ and $f_3$ (i.e. an error of one unit
in the determination of $N_1$, $N_2$ and $N_3$) leads respectively
to an error of $3.7 \cdot 10^{-11}$ s, $3.4 \cdot 10^{-11}$ s and
$ 1.0 \cdot 10^{-12}$ s on the clock desynchronisation,
significantly larger than the uncertainties aimed at.

The carrier phase ambiguity $N_i$ on the $f_i$ phase observable is
evaluated by combining the phase and code measurements and fixing
$N_i$ to the nearest integer. These unknown numbers of periods
added to phase measurements stay constant for continuous data ie.
as long as the signal acquisition is kept. This means we obtain
during a pass several evaluations of the same $N_i$ and then we
can average the noise on them to get a more accurate value.
Consequently correct cycle slip identification depends strongly on
the observable noise types and levels.

In order to express the phase ambiguities as a combination of
observables, we have to use a number of hypothesis. First the path
of code and phase measurements tagged with the same coordinate
time and at the same frequency are identical. We neglect the
change on geometric delays due to the opposite effect of
ionosphere on code and phase propagations. Moreover we assume all
downwards signals received at the same coordinate time by the
ground station have the same path through the atmosphere. Finally
the part of ionosphere crossed by upwards and downwards signals
are supposed identical for combined observables (i.e. for signals
within a few ms of each other).

When forming the difference between phase and code observables
measured at the same coordinate time and at the same frequency, we
cancel all the propagation terms except the ionospheric delay (see
equations (\ref{CodeC1}), (\ref{ExpressionT12}) and
(\ref{PhasePhi1})). Actually the ionospheric term depends on the
signal frequency
 and on the measurement nature, code or
carrier phase. The leading term is proportional to the inverse of
the square of the considered frequency. In addition, for the same
frequency, the effect of ionosphere on code is at first order the
opposite effect of ionosphere on carrier phase. The error on the
ionospheric delay comes from the error on the TEC determination
and is dominated by its leading term. Then the combination of
equations (\ref{CodeC1}) and (\ref{PhasePhi1}) for the ambiguity
$N_1$ on the first frequency can be written as :

\begin{equation}\label{ExpressionAmbiguityF1}
\frac{N_1}{f_1} = \Delta \tau_\phi^{s} \bigl(\tau^{s}(t_2)\bigl) -
\Delta \tau^{s} \bigl(\tau^{s}(t_2)\bigl) - 2 \cdot
\frac{40.308}{c f_1^2} TEC.
\end{equation}

The same kind of equation can be written for $f_2$ using code and
phase observables, $\Delta \tau^{g} \bigl(\tau^{g}(t_4)\bigl)$ and
$\Delta \tau_\phi^{g} \bigl(\tau^{g}(t_4)\bigl)$, or for the third
frequency $f_3$ with $\Delta \tau^{g}\bigl(\tau^{g}(t_6)\bigl)$
and $\Delta \tau_\phi^{g} \bigl(\tau^{g}(t_6)\bigl)$. They relate
the phase ambiguity $N_i$ with the difference between phase and
code observables and twice the ionospheric delay at the considered
frequency. Consequently the error on the phase ambiguity $\delta
N_i$ depends on the code and phase measurement noises,
respectively $\delta C_i$ and $\delta \Phi_i$, and on the error on
the TEC evaluation. The latter is determined with the code
measurements from $f_2$ and $f_3$ signals \cite{Duchayne}: 


\begin{equation}
\frac{40.308}{c} TEC = \frac{f_2^2 f_3^2}{f_2^2 - f_3^2}
\Biggl(\Delta \tau^{g} \bigl(\tau^{g}(t_4)\bigl) - \Delta \tau^{g}
\bigl(\tau^{g}(t_6)\bigl) \Biggl),
\label{TECFctCodeFrequence2&3Code}
\end{equation}


The Total Electron Content is evaluated with this formula and the
error on its determination is dominated by $f_2$ and $f_3$ code
measurement error. Hence we obtain, for each frequency $f_i$, the
dependence of the error on the phase ambiguity $\delta N_i$ on the
uncertainty in the observables by combining equations
(\ref{ExpressionAmbiguityF1}) and
(\ref{TECFctCodeFrequence2&3Code}):

\begin{equation}
\text{for $f_1$ : }\frac{\delta N_1}{f_1} = \delta \Phi_1-\delta
C_1 - 5.7 \cdot 10^{-2} (\delta C_2 - \delta C_3) ,
\label{AmbiguitesEtape1}
\end{equation}

\begin{equation}
\text{for $f_2$ : }\frac{\delta N_2}{f_2} = \delta \Phi_2-\delta
C_2 - 4.8 \cdot 10^{-2} (\delta C_2 - \delta C_3) ,
\label{AmbiguitesEtape2}
\end{equation}

\begin{equation}
\text{for $f_3$ : }\frac{\delta N_3}{f_3} = \delta \Phi_3-\delta
C_3 - 2.04 \cdot
 (\delta C_2 - \delta C_3) .\label{AmbiguitesEtape3}
\end{equation}

In the last equation (\ref{AmbiguitesEtape3}), two terms involve
the $f_3$ code noise $\delta C_3$. They are perfectly correlated
and partially cancel each other.

The noise model is evaluated with the latest results from the
engineering model \cite{Schaefer}. These results correspond to
code and phase measurements performed on a $f_1$ signal for a
received power of -95 dBm (this S/N ratio is related to a zenithal
position of the ISS over the ground station at about 350 km). They
show respectively a standard deviation of $1 \cdot 10^{-12}$~s and
$1  \cdot 10^{-13}$~s for the code and the phase measurements.
This 1 s noise will be used in this section to estimate the
statistics of cycle slips.

For this purpose, we assume the noise on $f_2$ approximatively the
same. On the contrary, the noise on $f_3$ is assumed to be 7
($\simeq f_1/f_3$) times larger on the carrier phase and 100
(ratio of chip rates) times larger on the code. In the following
table (Tab. \ref{TableauBruitWN}), we give the $2\sigma$
uncertainty (at 1~s points with no averaging) of each term
appearing in the equations (\ref{AmbiguitesEtape1}),
(\ref{AmbiguitesEtape2}) and (\ref{AmbiguitesEtape3}) . These
noise levels have to be compared with the half of the period
$1/f_i$ due to the fixing of the ambiguity $N_i$ to the nearest
integer : if the $2\sigma$ noise level is equal to half the
period, there is a 5 \% chance of a cycle slip.

\begin{table*}
\centering \caption{95 \% confidence ($2\sigma$ uncertainty)  of
terms appearing in the phase ambiguity resolution according to
equations (\ref{AmbiguitesEtape1}), (\ref{AmbiguitesEtape2}) and
(\ref{AmbiguitesEtape3}) for an input power of -95 dBm}
\begin{tabular}{|c|c|c|c|}
  \hline
 95 \% confidence ($2\sigma$ uncertainty) & $f_1$ &  $f_2$ &  $f_3$ \\
   \hline
  Noise on code (s)
 & $2  \cdot 10^{-12}$ &  $2  \cdot 10^{-12}$ &  $2  \cdot 10^{-10}$ \\
  Noise on carrier phase (s) & $2  \cdot 10^{-13}$ &  $2  \cdot 10^{-13}$  & $1.4  \cdot 10^{-12}$ \\
  Ionospheric error due to the noise on $f_3$ code (s)
 & $1.1  \cdot 10^{-11}$ &  $9.6  \cdot 10^{-12}$  & $2 \cdot 2  \cdot 10^{-10}$ \\
  Signal period $1/f_i$ (s)
 & $7.4 \cdot 10^{-11}$  & $6.7 \cdot 10^{-11}$ & $4.5 \cdot 10^{-10}$ \\
  \hline
\end{tabular}
  \label{TableauBruitWN}
  \end{table*}

We note that at all frequencies the cycle slip probability is less
than 5 \% - remember that on $f_3$ the code noise and half the
ionosphere noise cancel (see equation (\ref{AmbiguitesEtape3})) -
even when using individual 1 s measurements (no averaging). The
probability can be further reduced by averaging the measurements
of a continuous observation.

However two facts will disturb the cycle slip identification.
First the received power does not remain at -95 dBm but decreases
when the distance between the two stations increases. As an
example, when the space station is at a ten degrees elevation, the
received signal power is -115 dBm. From the tests of the
engineering model, this corresponds to an eight fold increase of
the code noise. In these worst conditions, the direct method of
ambiguity resolution fails. One solution is to carry out a
weighted average with all 1 s measurements of a given continuous
pass. Indeed, the noise level of each measurement will depend on
the incoming power (varying as a function of elevation). The
dependence of the noise on power has been measured so the
weighting can be preformed as a function of the locally measured
power level.

Secondly the measurement noise is not perfectly white. Some
flicker noise damages the averaging and the ambiguity removal,
 and even averaging over the complete pass is insufficient due to the
presence of flicker noise. We need to search for observable
combinations to increase the ambiguity resolution success rate.


The figures of Tab. \ref{TableauBruitWN} show that the leading
error terms come from the code noise on the third frequency. Our
aim is to get rid of the code measurement of $f_3$, or at least to
reduce its impact on the final statistics.

We start by noting that, according to the noise levels in Tab.
\ref{TableauBruitWN}, the ambiguity resolution for the Ku signals
($f_1$ and $f_2$) is more likely to be successful than for the
third frequency. Then, if we suppose these ambiguity resolutions
are correctly performed (this is likely the case, see section
\ref{TestsNoise}), these considerations bring two new methods for
the $f_3$ signal ambiguity resolution. On the one hand, the
combination of $f_2$ and $f_3$ phase measurements gives a new
expression of $N_3$ which does not depend on the $f_3$ code
measurement :

\begin{equation}
\begin{split}
 \frac{N_3}{f_3} &= \Delta \tau_\phi^{g} \bigl(\tau^{g}(t_6)\bigl) -
\Delta \tau_\phi^{g} \bigl(\tau^{g}(t_4)\bigl)\\
& + \frac{N_2}{f_2} + \frac{f_2^2 - f_3^2}{f_2^2
f_3^2}\frac{40.308}{c} TEC .\\
\label{TECFctPhaseF2&F3} \end{split}
\end{equation}

In the previous equation, the Total Electron Content is evaluated
by the combination of $f_2$ phase and code measurements~:

\begin{equation}
 \frac{40.308}{c} TEC =
\frac{f_2^2}{2} \Biggl( \Delta \tau_\phi^{g}
\bigl(\tau^{g}(t_4)\bigl) - \Delta \tau^{g}
\bigl(\tau^{g}(t_4)\bigl)- \frac{N_2}{f_2} \Biggl) .
\label{TECFctCode&PhaseF2}
\end{equation}

It allows a new evaluation of the TEC whose error depends only on
$\delta C_2$ and $\delta \Phi_2$ and which  can replace the
equation (\ref{TECFctCodeFrequence2&3Code}). Inserting equation
(\ref{TECFctCode&PhaseF2}) into (\ref{TECFctPhaseF2&F3}) leads to

\begin{equation}
\frac{\delta N_3}{f_3} = \delta \Phi_3 - 22.72 \cdot \delta \Phi_2
+ 21.72 \cdot \delta C_2 . \label{Methode1}
\end{equation}

On the other hand, the second method is to substitute
(\ref{TECFctCodeFrequence2&3Code}) into
(\ref{ExpressionAmbiguityF1}) (but written for $f_3$) for half the
ionospheric term and using (\ref{TECFctCode&PhaseF2}) for the
other half. The error from the $f_3$ code measurement almost
cancels, and the error on $N_3$ is :

\begin{equation}
\frac{\delta N_3}{f_3} = \delta\Phi_3 + 2.3 \cdot 10^{-2} \cdot
\delta C_3 - 22.22 \cdot \delta\Phi_2  + 21.20 \cdot \delta C_2.
\label{Methode2}
\end{equation}


Equations (\ref{Methode1}) and (\ref{Methode2}) reduce
the error from the code measurements from $f_3$ and then increase
the success rate. The choice between one of these methods depends
on the real mission noise levels. With the noise measured in the
engineering model tests, they both  approximatively give a five
fold gain on the uncertainty of $\delta N_3$.

In conclusion, the statistics of successful carrier phase
ambiguity resolution strongly depend on the pass characteristics
and its relation with the noise levels. These kind of realistic
received noises are studied in the following section
\ref{TestsNoise}.

\section{ Tests with expected noise}
\label{TestsNoise}

In this section, we investigate the error on the phase ambiguity
resolution from realistic measurement noises and try to reduce the
probability of cycle slips on the ambiguity identification. For an
ISS pass over a ground station -ie. for time duration less than
600~s-, the noise measured on the Engineering model can be
modelled in a simplified from by its temporal Allan deviation. We
use respectively for code and phase measurements at -95~dBm
$\bigl(\sigma_x^c (\tau)\bigl)^2 = (1\cdot10^{-12}
\tau^{-\frac{1}{2}})^2 +(2\cdot10^{-13})^2~s$ and
$\bigl(\sigma_x^\phi (\tau)\bigl)^2 = (1\cdot10^{-13}
\tau^{-\frac{1}{2}})^2 +(7\cdot10^{-14})^2~s$, where $\tau$ is the
integration time \cite{Schaefer}.

\begin{figure}[htbp]
\includegraphics[height=6cm]{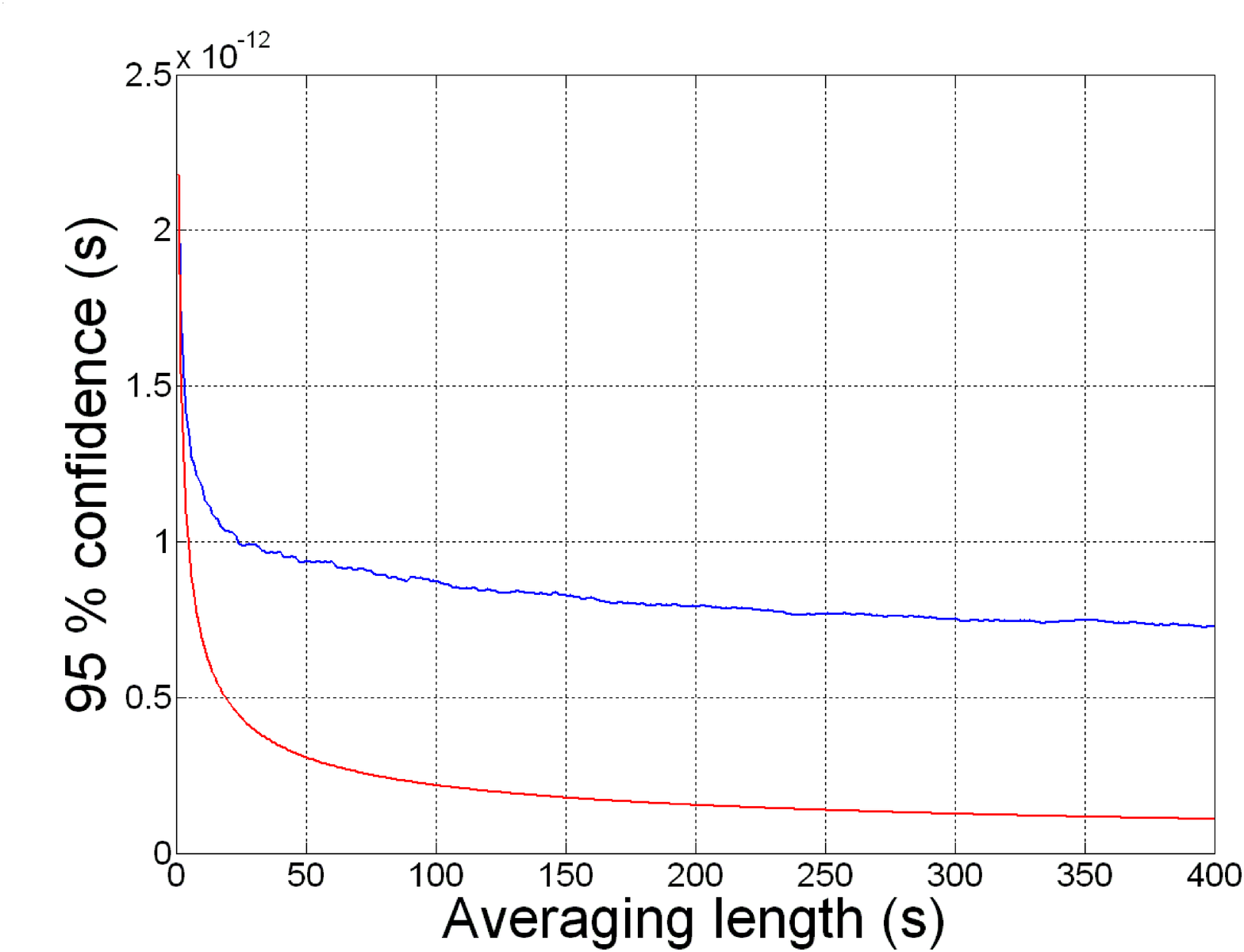}
\caption{95 \% confidence of averaging of white noise (red) and
$f_1$ code measured noise at -95 dBm (blue). Both have the same
deviation at one second.} \label{ComparaisonAveragingWNandFN}
\end{figure}

The measured noise is composed of white noise and flicker noise.
The former is averaged rapidly as the square root of time. Then
the noise average is limited by the flicker noise. On figure
(\ref{ComparaisonAveragingWNandFN}), we compare the averaging of
the measured noise on $f_1$ code at -95 dBm, with the averaging of
a pure white noise with the same level at one second ($\sigma = 1
\cdot 10^{-12}$~s).

During a pass of the International Space Station over a ground
station, the signal over noise ratio strongly varies and its
received signal power can decrease by over 20 dBm. This ratio
actually depends on the distance between the two stations, the
elevation of ISS and others parameters (antenna orientation,
weather conditions, etc ...). A model provided by EADS-Astrium
gives power evolution with distance and elevation. The engineering
model measurements of the noise dependence on power carried out at
TimeTech \cite{Schaefer} then allow us to derive the dependence of
the noise on elevation and distance, i.e. the variation of the
noise during a given pass.

 We perform Monte Carlo simulations
to estimate the probabilities of a cycle slip and its
uncertainties (95 \% confidence). In fact, we create code and
phase signals with the same noise properties as the measured noise
on the engineering model for the frequency $f_1$. The $f_2$ and
$f_3$ noise for code and phase measurements are similarly
generated taking into account chip rate or frequency ratios. Then
these noises are modulated as a function of the elevation and
distance for a given pass. In a pessimistic approach we suppose
the two kinds of noise (white and flicker)  on code and phase
measurements are affected in the same way by the received power.
 Hence we can calculate the
95 \% confidence of the ambiguity error in accordance with
equations (\ref{AmbiguitesEtape1}), (\ref{AmbiguitesEtape2}),
(\ref{AmbiguitesEtape3}), (\ref{Methode1}) and (\ref{Methode2}).

The noise on the ambiguity identification is expressed in seconds
: as this noise has to be compared with the signal period, we
express it in term of a fraction $\delta N_i$ of the considered
signal cycle. We average over a number of 1 s points to decrease
the uncertainty on $N_i$. Then, if $\delta N_i$ after averaging
exceeds half a cycle with more than 5 \% probability, we consider
that the ambiguity resolution has failed as $N_i$ is fixed to the
wrong number. As an improvement over simple averaging, we also
consider weighted averaging, where the weights are determined by
the noise level of each point (depending on received power, ie.
elevation and distance).

In the following we study the time transfer between the
International Space Station and a ground station based in
Toulouse, France $(43^{o} 36' N, 1^{o} 26' E)$ to simulate
realistic noise levels. Actually, this station has been chosen as
the master ground station of the ACES mission. Then we choose an
ISS trajectory corresponding to one pass over the ground station.
To this purpose, we consider an ephemeris of ISS corresponding to
the $20^{th}$ of May, 2005. However in order to simulate several
passes with different elevation higher than four degrees, we
slightly shift the time origin of the ground station trajectory.
This way we obtain numerous passes characterized by their reached
maximum elevation and whose duration is between 350 and 600~s.

\begin{figure}[htbp]
\includegraphics[height=6cm]{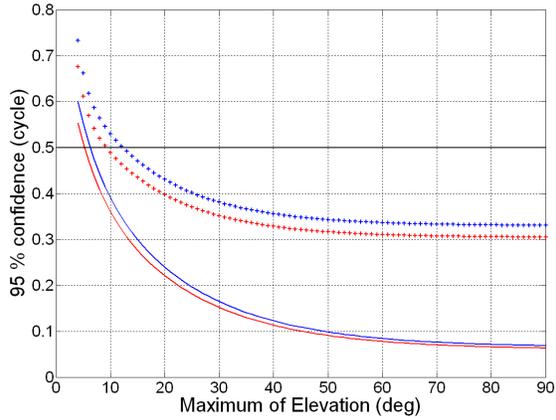}
\caption{95 \% confidence of the phase ambiguity resolution error
as a function of the maximal elevation of the pass. The four
curves show this dependence for $f_1$ (blue) and $f_2$ (red)
signals using arithmetic (cross-dashed) or weighted (solid)
averaging in accordance with equations (\ref{AmbiguitesEtape1})
and (\ref{AmbiguitesEtape2}).} \label{ComparisonN1N2WeightedOrNot}
\end{figure}

Figure (\ref{ComparisonN1N2WeightedOrNot}) shows the uncertainties
(95 \% confidence) of $\delta N_1$ and $\delta N_2$ as a function
of the maximal elevation reached by the space station during the
pass. These uncertainties are calculated with and without a
weighted averaging over a complete pass ie. data are acquired
continuously during the pass. It allows to assess the effect of
weighting on $N_1$ and $N_2$ final estimation.

Weighted averaging leads to a gain on the success rate and
particularly with high elevations : at ninety degrees elevation,
the statistics are almost divided by a factor six between the two
averaging methods. In fact, at lower elevations, the noise level
on the observables does not sensibly change during the pass, and
in this case the weighting becomes inefficient. We note that above
six degrees maximum elevation, the phase ambiguity is correctly
determined (95 \% confidence).

Finally we investigate the impact of averaging (arithmetic or
weighted) and calculation (in accordance with equations
(\ref{AmbiguitesEtape3}), (\ref{Methode1}) or (\ref{Methode2}))
methods on the uncertainty on the $f_3$ phase ambiguity $N_3$. For
the noise considered in this paper, the methods coming from
equations (\ref{Methode1}) and (\ref{Methode2}) give
approximatively identical results.

\begin{figure}[htbp]
\includegraphics[height=6cm]{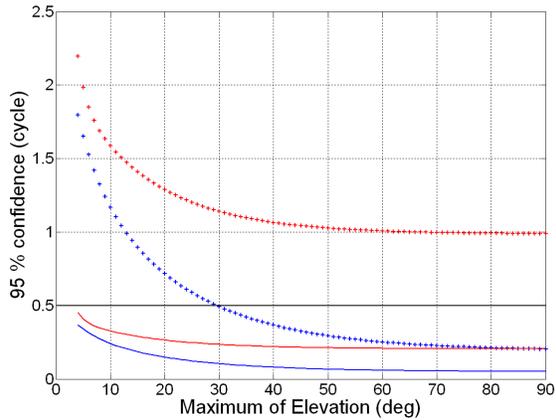}
\caption{95 \% confidence of the phase ambiguity resolution error
for the $f_3$ signal as a function of the maximal elevation of the
pass. The four curves show this evolution whether we consider
first equation (\ref{AmbiguitesEtape3}) (cross-dashed) or proposed
method  $\bigl($(\ref{Methode1}) or (\ref{Methode2})$\bigl)$
(solid) and whether arithmetic (red) or weighted (blue) averaging
is performed.} \label{ComparisonMethods1&2WeightedOrNot}
\end{figure}

Figure \ref{ComparisonMethods1&2WeightedOrNot} depicts the
different combination of approaches concerning the averaging and
the calculation.  Contrary to weighting, the observable
combinations given by equation (\ref{Methode1}) or
(\ref{Methode2}) brings a factor five  gain for all elevations
(see the two lowest curves on figure
\ref{ComparisonMethods1&2WeightedOrNot}). We also remark that the
proposed measurement combinations (\ref{Methode1}) or
(\ref{Methode2}) are essential in the cycle slip detection on the
third frequency, and thus for the mission objectives.

However the obtained probability of successful ambiguity
resolution on $f_3$  needs correct identification of the Ku cycle
slips ($f_1$ and $f_2$). Consequently, an unbiased estimation of
the scientific products is not limited anymore by the $f_3$ phase
ambiguity measurement but by the two first frequencies. But in any
case, using weighted averaging and methods (\ref{Methode1}) and
(\ref{Methode2}) for $f_3$, the cycle ambiguity resolution has
less than 1 \% probability to fail for any uninterrupted satellite
pass with a maximum elevation over four degrees if we suppose that
$f_1$ and $f_2$ phase ambiguities have been correctly solved.

\section{ Discussion and conclusion }

The models of the time transfer of the ACES mission and the
associated relativistic effects were investigated at the 0.1
picosecond level. In addition, the measurements were related to
all physical effects appearing during the station communication.

Then these models were applied to develop an end-to-end
simulation. In fact, two softwares were implemented : a simulation
which creates the raw measurements of the mission and allows
addition of different kinds of perturbations, and an algorithm
which uses these raw measurements to extract scientific products.

Furthermore the MWL phase ambiguity resolution was tackled and
methods were studied to reduce the failure rate of this process.
The cycle slip removal faced two difficulties, the presence of
Flicker noise which can not be averaged, and the evolution of the
signal over noise ratio during the pass. Considering a model for
the power evolution with distance and elevation, we simulated
noises complying with experimental code and phase measurements on
the first frequency $f_1$. Then the $f_2$ and $f_3$ code and phase
noises were deduced assuming they are related to $f_1$ code and
phase noise with respectively chip rate and frequency ratios. In a
pessimistic approach, we supposed all components of noise (white
and Flicker) are similarly modulated by the S/N ratio.

With Monte Carlo simulations, we obtained the most realistic
estimates of the resulting performance for uninterrupted passes.
The success rate is now limited by the ambiguity resolution of the
Ku-band signals. In fact the high probability of successful
resolution on the third frequency is based on the hypothesis that
there is no cycle slip on $f_1$ and $f_2$. Their resolutions are
limited by the code noise from frequency $f_3$. The failed
resolution proportion for Ku-band signals is less than 5 \% for
passes with maximum elevation over six degrees. Consequently, in
this case, the phase ambiguity resolution is successful for all
three frequencies with more than 94 \% probability. An increase of
the S-band signal chip rate, from $10^{6} s^{-1}$ to $5 \cdot
10^{6} s^{-1}$, would strongly improve the Ku-band signal phase
ambiguity as it will approximatively divide $f_3$ code noise by a
factor 5.

On one hand, these statistics are based on a pessimistic
hypothesis of noise variation with received power. We assume
indeed all kind of noise in the measurement are similarly
modulated by the power evolution whereas it is likely that the
Flicker noise is not or less affected. On the other hand, we have
optimistically considered only completed uninterrupted passes over
the ground station. In a close future, we will investigate the
phase ambiguity resolution for incomplete passes (loss of lock
during the pass) and other perturbations affecting the raw
measurements during a pass (eg. temperature variations).

\

\textbf{Acknowledgements}

\

 The research project is supported by the
French space agency CNES, ESA and EADS-Astrium through Lo\"\i c
Duchayne's research scholarship $N^{o}$ 05/0812.

\


\end{document}